\documentclass[reprint,aps,prl,superscriptaddress,longbibliography,linenumbers]{revtex4-2}

\pdfoutput=1 

\usepackage{hyperref}
\hypersetup{colorlinks=true, citecolor=black, urlcolor=black, linkcolor=black}

\usepackage{graphicx}% Include figure files
\usepackage{dcolumn}% Align table columns on decimal point
\usepackage{braket}
\usepackage{bm}
\usepackage{color}
\usepackage{amsmath}

\begin{document}
\nolinenumbers
%\preprint{}

\title{Atom Camera: Super-resolution scanning microscope of a light pattern with a single ultracold atom}


\author{T. Tomita}
\affiliation{Institute for Molecular Science, National Institutes of Natural Sciences, Okazaki, Japan}
\affiliation{SOKENDAI (The Graduate University for Advanced Studies), Okazaki, Japan}

\author{Y. T. Chew}
\affiliation{Institute for Molecular Science, National Institutes of Natural Sciences, Okazaki, Japan}
%\affiliation{SOKENDAI (The Graduate University for Advanced Studies), Okazaki 444-8585, Japan}

\author{R. Villela}
\affiliation{Institute for Molecular Science, National Institutes of Natural Sciences, Okazaki, Japan}
\affiliation{SOKENDAI (The Graduate University for Advanced Studies), Okazaki, Japan}

\author{T. P. Mahesh}
\affiliation{Institute for Molecular Science, National Institutes of Natural Sciences, Okazaki, Japan}
\affiliation{SOKENDAI (The Graduate University for Advanced Studies), Okazaki, Japan}

\author{H. Sakai}
\affiliation{Central Research Laboratory, Hamamatsu Photonics K.K, Hamamatsu, Japan}

\author{K. Nishimura}
\affiliation{Central Research Laboratory, Hamamatsu Photonics K.K, Hamamatsu, Japan}

\author{T. Ando}
\affiliation{Central Research Laboratory, Hamamatsu Photonics K.K, Hamamatsu, Japan}

\author{S. de L\'es\'eleuc}
%\email[]{sylvain@ims.ac.jp}
\affiliation{Institute for Molecular Science, National Institutes of Natural Sciences, Okazaki, Japan}
%\affiliation{SOKENDAI (The Graduate University for Advanced Studies), Okazaki 444-8585, Japan}
\affiliation{RIKEN Center for Quantum Computing (RQC), Wako, Japan}

\author{K. Ohmori}
%\email[]{ohmori@ims.ac.jp}
\affiliation{Institute for Molecular Science, National Institutes of Natural Sciences, Okazaki, Japan}
\affiliation{SOKENDAI (The Graduate University for Advanced Studies), Okazaki, Japan}

%\thanks{}

\date{\today}

\begin{abstract}
Sub-micrometer scale light patterns play a pivotal role in various fields, including biology~\cite{Dholakia2011}, biophysics~\cite{Bustamante2021}, and AMO physics \cite{Browaeys2020,Kaufman2021}. High-resolution, \textit{in situ} observation of light profiles is essential for their design and application. However, current methods are constrained by limited spatial resolution and sensitivity. Additionally, no existing techniques allow for super-resolution imaging of the polarization profile, which is critical for precise control of atomic and molecular quantum states. Here, we present an \textit{atom camera} technique for \textit{in situ} imaging of light patterns with a single ultracold atom held by an optical tweezers as a probe. By scanning the atom's position in steps of sub-micrometers and detecting the energy shift on the spin states, we reconstruct high-resolution 2D images of the light field. Leveraging the extraordinarily long coherence time and polarization-sensitive transitions in the spin structure of the atom, we achieve highly sensitive imaging both for intensity and polarization. We demonstrate this technique by characterizing the polarization in a tightly-focused beam, observing its unique non-trivial profile for the first time. The spatial resolution is limited only by the uncertainty of the atom's position, which we suppress down to the level of quantum fluctuations ($\sim$ 25 nm) in the tweezers' ground state. We thus obtain far better resolution than the optical diffraction limit, as well as than the previous ones obtained with a thermal atom fluctuating in the trap~\cite{Deist2022,Mitchell2022}. This method enables the analysis and design of submicron-scale light patterns, providing a powerful tool for applications requiring precise light manipulation.

\end{abstract}

\maketitle

The generation of light fields using tightly-focused laser beams has become increasingly important in a variety of fields. Optical tweezers, for instance, are used to capture and control target specimens in biology ~\cite{Dholakia2011}, single-molecule biophysics \cite{Bustamante2021}, and to manipulate the quantum state of atoms and molecules \cite{Browaeys2020,Kaufman2021}. Not only simple Gaussian beams, but also spatially structured light is harnessed \cite{Yang2021}, which allows for advanced fluorescence imaging \cite{Hell1994,Willig2006,Mudry2012} and investigation of nano-plasmonics \cite{Failla2006,Banzer2010}. In the context of cold-atom experiments, there is a growing interest for the generation of arbitrarily-shaped far-field patterns \cite{Weitenberg2011,Krinner2015,Zupancic2016,Saffman_superGauss,Barredo2020,Chew2022,Chew2024}, including beams with sub-wavelength super-oscillatory features~\cite{Wilkowski2022,Zheludev2022review}, as well as for exploiting the near-field close to nano-photonics devices~\cite{Chang2009,Kimble2016,Dordevic2021,Menon2023}. In the field of quantum computations with atoms~\cite{Bluvstein2024,Radnaev2024}, the quality of the quantum states is affected by the intensity and the polarization of the light used to control the atoms~\cite{Thompson2013, Unnikrishnan2024, Chew2024}.

In such experiments, \textit{in situ} observation of the light field is essential. An external diagnostic is often not adequate because the light field would not propagate (optical lattices, near-field features, ...) or some effects are only present upon tight-focusing with a high-NA objective lens (non-paraxial effects) which is not compatible with the finite resolution of a camera, and overall because the additional optical path could introduce aberrations and limit the resolution. \textit{In situ} super-resolution can be achieved by scanning a nanoscopic probe interacting with the light field~\cite{Rotenberg2014,Long2020}, where the probe can even be a single atom or ion~\cite{Guthohrlein2001, Steiner2013, Lee2014, Deist2022, Mitchell2022}.

Here, we report on an \textit{atom camera} method for imaging the intensity and polarization of a light pattern using a single ultracold $^{87}$Rb atom trapped in an optical tweezers. By scanning the atom's position with sub-micrometer steps and measuring the perturbation made by the light to the valence electron energy, we reconstruct a 2D image of the light pattern. 
The spatial resolution of this scanning microscope is ultimately limited only by the quantum fluctuation of the atom's position ($\sim 25$ nm), i.e., the size of the atomic wavefunction in the tweezers' ground state, an order of magnitude better than the diffraction limit at optical wavelengths and the ones from earlier pioneering works using a thermal atom fluctuating in the trap by several $100$s of nanometers~\cite{Deist2022, Mitchell2022}.

\begin{figure}
	\includegraphics[width=\linewidth]{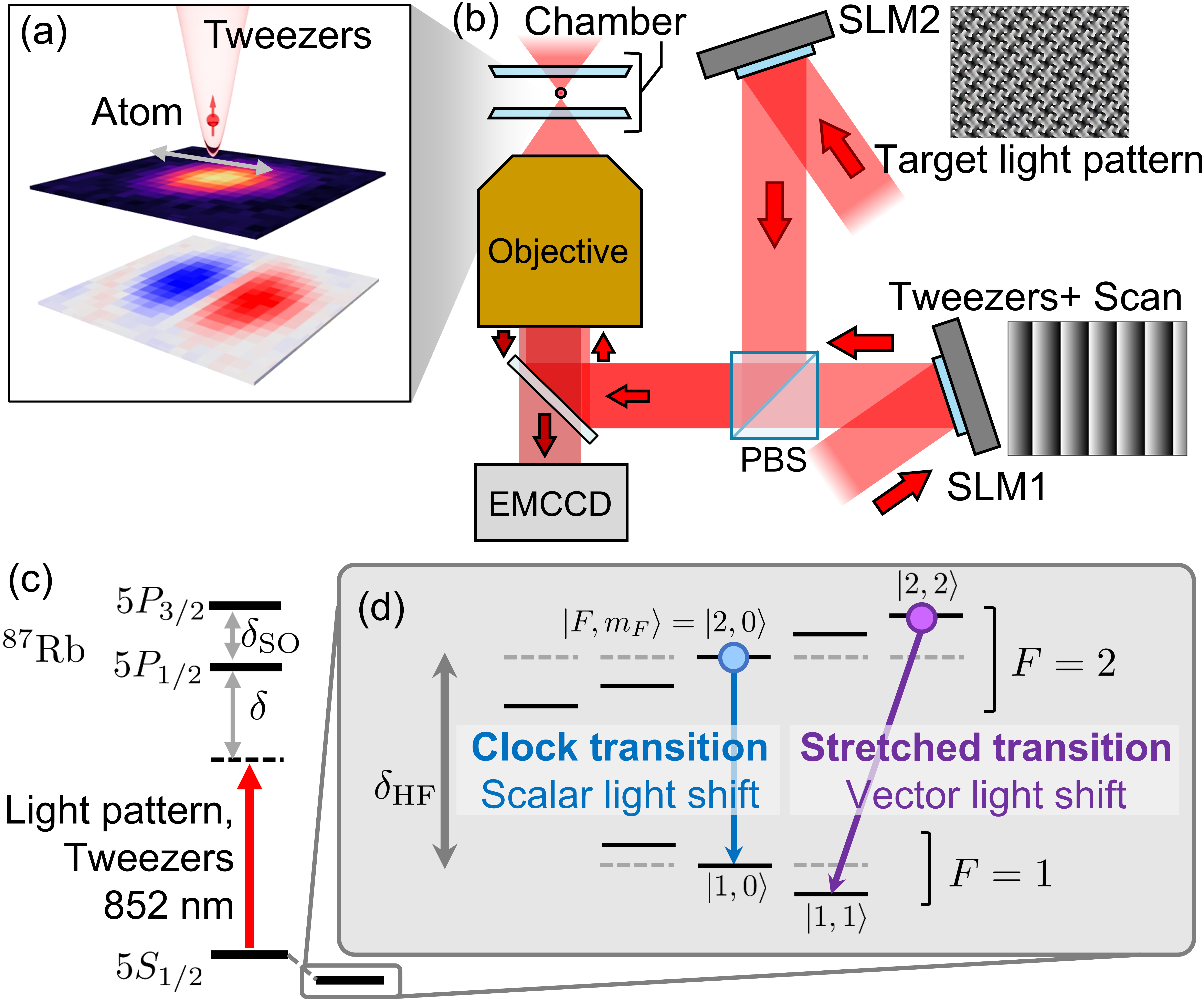}
	\caption{\label{fig1} \textbf{A single ultra-cold atom probe.} (a) Schematic of a single  $^{87}$Rb atom trapped in the motional ground-state of an optical tweezers and scanned over a light pattern. The top (bottom) image represents the measured light intensity (polarization). (b) Optical setup consisting of a high-NA objective lens and two spatial light modulators (SLM). The first SLM (SLM1) generates an optical tweezers trapping the atom (the probe) whose position can be finely scanned by adjusting the grating pattern on the SLM. The second SLM (SLM2) generates an arbitrary target light pattern. The two beams have orthogonal polarizations and are combined on a polarization beam splitter. (c) Relevant energy levels: electronic ground-state $5S_{1/2}$ with its hyperfine spin structure split by $\delta_{\rm HF} = 6.8$~GHz and first excited orbitals $5P_{1/2}$ and $5P_{3/2}$ split by the spin-orbit coupling $\delta_{\rm SO} = 7$~THz. Both tweezers and light pattern are at 852 nm, red-detuned by $\delta = -30$~THz from the excited state, such that they light-shift the $5S_{1/2}$ states to lower energy. (d) Zoom in on the hyperfine spin structure used to probe the intensity and polarization of the light. The clock transition $\ket{2,0} \leftrightarrow \ket{1,0}$ is sensitive to the intensity of the target light due to the scalar differential light shift (DLS), while the stretched transition $\ket{2,2} \leftrightarrow \ket{1,1}$ is sensitive to the polarization of the target light through the vector DLS. 
	}
\end{figure}

A second novelty of our method is to sense the light field with the electron's spin degree of freedom (d.o.f.) of the atom [Fig.~\ref{fig1}(a)], which can be coherently manipulated, instead of using the electron's orbital d.o.f. (i.e., an optical transition between electronic orbital states)~\cite{Guthohrlein2001, Steiner2013, Lee2014, Deist2022, Mitchell2022}. The light induces a perturbative shift of the valence electron energy in the ground-state orbital: the light-shift (LS). Rather than directly measuring the orbital LS, we interrogate the much weaker differential light-shift (DLS) of the two hyperfine spin configurations of the ground-state ($F=1$ and $F=2$ in $^{87}$Rb), originating from the slight difference of their resonance frequencies due to the hyperfine splitting. While the DLS is 3 to 4 orders of magnitude smaller than the orbital LS, we nevertheless obtain a sensitivity improved by an order of magnitude thanks to the extraordinarily long coherence time of the hyperfine states that can reach up to a second~\cite{Zhan2020}, to be compared to the 26~ns lifetime of the orbital excited state used to interrogate the LS in previous works~\cite{Brantut2008, Deist2022, Mitchell2022}.

Using this electron's spin d.o.f. allows us to demonstrate that the single atom can be used as a sensitive probe for the light polarization. By interrogating a magnetic-field insensitive hyperfine transition, we probe the scalar DLS proportional only to the light intensity, as discussed in the previous paragraph. If we instead use a magnetic-field sensitive transition, we then additionally probe the vector DLS proportional to the ellipticity of the light polarization. Intuitively, circularly polarized light - a rotating electric field - drives the electron cloud to rotate, perturbing the energy of the hyperfine spin states (the vector LS) through spin-orbit interaction. Using this technique, we were able to map the non-trivial polarization profile of a tightly-focused optical tweezers for the first time, to the best of our knowledge, and obtain results in agreement with vector diffraction theory~\cite{Inoue1952,Wolf1959,Richards1959}.

%%%%%%%%%%
%%%  Fig 1  %%%
%%%%%%%%%%

\section*{Results}

\subsection*{Atom-camera manual}

The cold-atom experimental setup has been described previously~\cite{Chew2022}. Briefly, we trap a single $^{87}$Rb atom in an optical tweezers obtained by focusing an 852 nm trapping beam with a high-NA objective (NA=0.75 in design). The optical tweezers' position is controlled by a hologram displayed on a first spatial light modulator [SLM1 in Fig.~\ref{fig1}(b)], while a second SLM generates a pattern of light, also at 852~nm, to be imaged with our \textit{atom camera}. Using Raman sideband cooling, we actively prepare the atom in the motional ground-state of the tweezers with an rms spread of the wavefunction of 25 nm setting the resolution limit of the scanning microscope. 

The 852~nm lasers shift the energy of the $5S$ ground orbital by $U$ (typically, $-h\times$ 10 MHz for the tweezers), which takes slightly different values depending on the hyperfine spin configuration $\ket{F,m_F}$ giving rise to a differential light-shift $\Delta U = U_{F=2} - U_{F=1}$. A first origin of this effect is the hyperfine splitting $\delta_{\rm HF} = 6.8$~GHz causing a scalar DLS $\Delta U_s$ (independent of light polarization and magnetic sublevels $m_F$) given by $\Delta U_s/U = \delta_{\rm HF}/|\delta|  \simeq 2 \times 10^{-4}$ for our laser detuning $\delta \simeq -30$~THz.  Secondly, a vector DLS appears from the combination of elliptically polarized light and the spin-orbit coupling in the excited 5$P$ orbital ($\delta_{\rm SO} = 7$~THz) amounting to $\Delta U_v/U = -\delta_{\rm SO}/3\delta \times C \Delta(g_F m_F) \simeq 0.1 C$, with $C$ the polarization ellipticity (see Methods) and $g_F$ the hyperfine $g$-factor. $\Delta (g_F m_F)$ is the difference of the product of $g_F m_F$ between two hyperfine spin states. As illustrated in Fig.~\ref{fig1}(d), the clock transition $\ket{2,0} \leftrightarrow \ket{1,0}$ is unaffected by the vector DLS and is shifted only by the scalar DLS on the order of a kHz. In contrast, the stretched transition $\ket{2,2} \leftrightarrow \ket{1,1}$ experiences a vector DLS that dominates the scalar contribution and is on the order of 10 kHz for $C = 0.01$ (almost purely linear polarization) and 1 MHz for $C = 1$ (purely circular) for the intensity of typical tweezers light.

\begin{figure}
	\includegraphics[width=\linewidth]{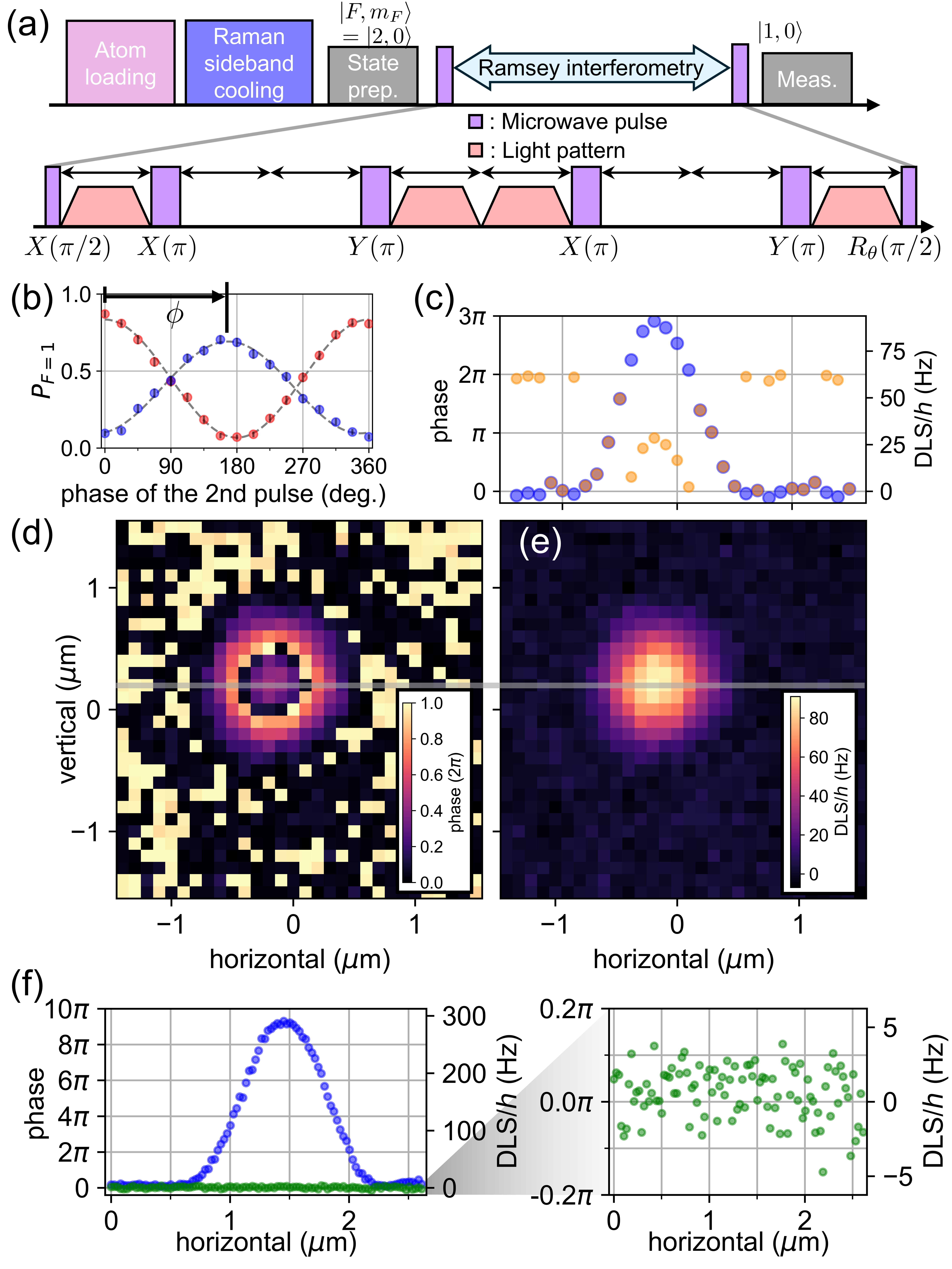}
	\caption{\label{fig2} \textbf{Atom camera working principle.} (a) Ramsey interferometer with a dynamical decoupling sequence (XY-4 for the scalar DLS measurement). $X(\varphi)$, $Y(\varphi)$ and $R_{\theta}(\varphi)$ represent $\varphi$-rotation pulse with microwave phase of 0$^{\circ}$, 90$^{\circ}$ and $\theta$, respectively. (b) A typical Ramsey fringe obtained while scanning the phase of the closing $\pi/2$-pulse and measuring the population in the $F=1$ state. The fringe with the target light pattern (blue) is shifted by $\phi$ from the reference fringe (red, with the pattern off). (c-e) The imaged light profile. (d) The 2D profile of the phase shift $\phi$ for a tightly-focused Gaussian beam. The phase shift is wrapped modulo 2$\pi$. The step size of the probe scan is 100 nm. (e) The image of the DLS after applying a phase unwrapping algorithm. The phase unwrapping procedure is seen in the 1D cut image (c), reconstructing the original intensity profile (blue) from the raw phase data (orange). The DLS are plotted as positive values for correspondence with the light intensity. (f) 1D scan image for the signal-to-noise ratio evaluation. We compare the data with (blue) and without (green) a target light pattern (right). The step size is reduced to 25 nm to avoid the large phase difference between the neighboring points. The noise level is purely determined by the quantum projection noise (left).}
\end{figure}

We measure the scalar and vector DLSs using the Ramsey interferometer shown Fig.~\ref{fig2} (a): we prepare the atom in a superposition of two hyperfine spin states and then superimpose the pattern to be imaged which shifts the Ramsey fringe, see Fig.~\ref{fig2}(b), by $\phi = \Delta U/h \times \tau_{\rm eff}$ with $\tau_{\rm eff}$ the effective interrogation time. To avoid the atom's position to be shifted and heated when switching on and off the light pattern (see Methods), which would result in blurring and distortion of the image, we always keep this pattern more than 10 times weaker than the optical tweezers (such that the atom displacement is smaller than the resolution limit). Consequently, the DLS from the pattern is hidden by the much stronger tweezers DLS, with the latter varying spatially (when scanning the tweezers position) and temporally (power drift). To remove this large, noisy, contribution and isolate the weak signal of interest, we insert a dynamical decoupling sequence (``XY-4" for the scalar DLS measurement~\cite{Gullion1990} and simple echo for the vector DLS measurement) in the Ramsey interferometer while switching on and off the pattern. It also helps to get higher sensitivity by extending the coherence time and thus the interrogation time.

%%%%%%%%%%
%%%  Fig 2  %%%
%%%%%%%%%%

We extract the phase from the Ramsey signal [fringe in Fig. 2 (b)]. The rms value of the signal without the target light is $2\pi \times 0.0275$ [Fig.~\ref{fig2}(f)] while the quantum projection noise (QPN) of phase estimation is $\Delta \phi_{\rm QPN} = 1/V\sqrt{N} \simeq 2\pi \times 0.0283$, achieving the QPN limit for the readout noise. Here, $V$ is the contrast of the Ramsey fringe and $N$ is the number of shots. The sensitivity of our atom camera to the light-shift $U$ becomes $\sim 34~{\rm kHz}/\sqrt{\rm Hz}$ at 852 nm ($\sim$ 10 s to get 10 kHz error, scaling as $|\delta|$), which could be further improved by increasing the interrogation time or the duty cycle (only 20 \% of the cycle is spent for Ramsey interferometry, the rest being preparation and measurement). We note that this sensitivity is already ten times better than when using the optical transition \cite{Deist2022}, demonstrating the advantage of coherently probing the electronic spin.

%%%%%%%%%%%%%%%%%%%%%%%%%%%%%%%%%%%%%%%%%%%%%%%%%%%%%%%%
\subsection{Intensity profile: Scalar light-shift measurement}

%%%%%%%%%%
%%%  Fig 3  %%%
%%%%%%%%%%
\begin{figure}
	\includegraphics[width=\linewidth]{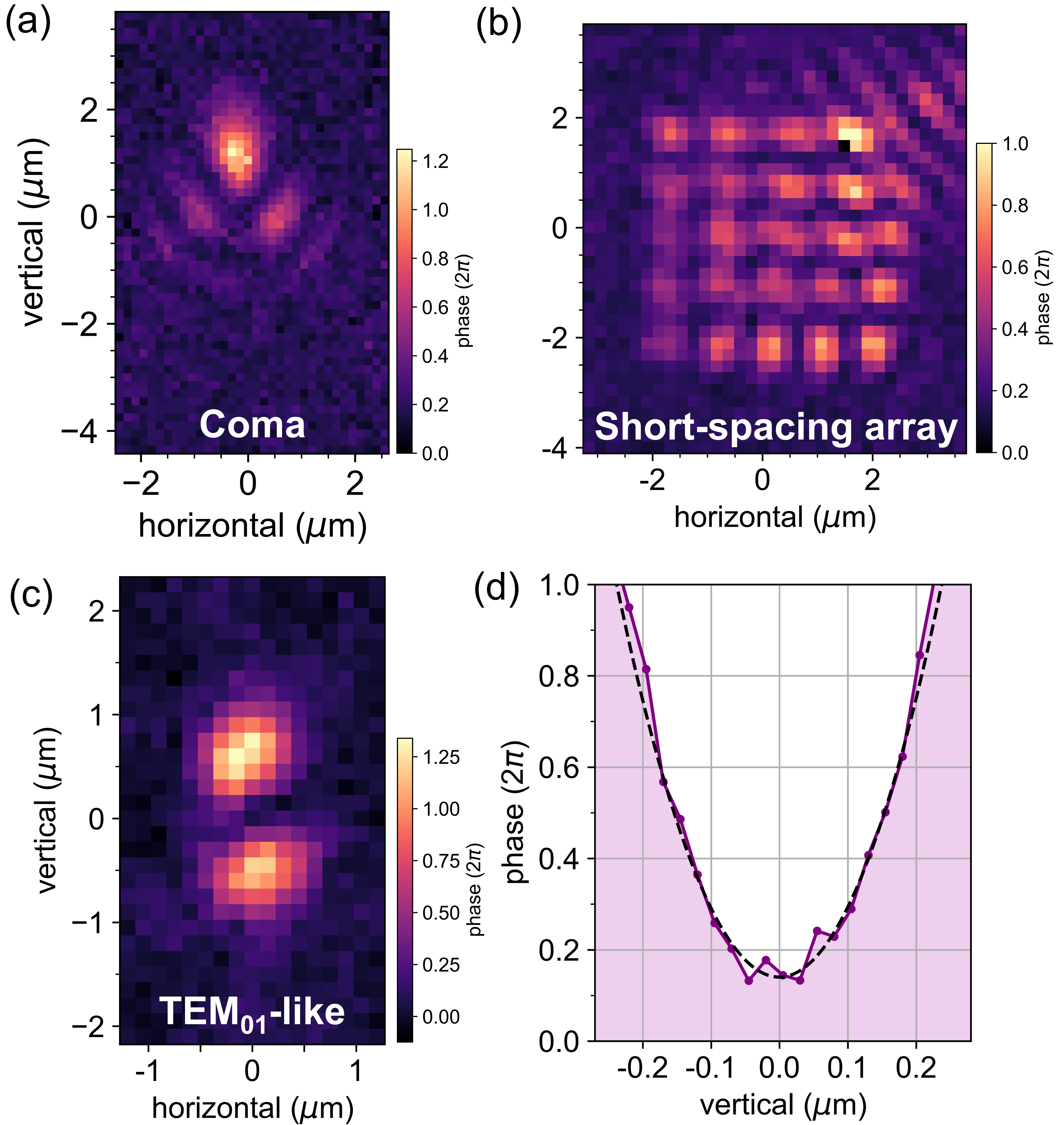}
	\caption{\label{fig3} \textbf{Gallery of \textit{in-situ} super-resolved images and the evaluation of the spatial resolution.} (a) a tightly-focused beam with strong coma, (b) a short-spacing array, (c) a TEM$_{01}$-like pattern. (d) The evaluation of the spatial resolution. The phase shift data at the center of the TEM$_{01}$-like pattern (purple) is fitted by a quadratic function (dotted line). }

\end{figure} 

We now demonstrate the full 2D reconstruction of the intensity profile of a pattern. First, we image a simple tightly-focused Gaussian beam, i.e., an optical tweezers. We measure the scalar DLS while scanning the single-atom probe with steps of 100 nm, calibrated with an optical ruler method~\cite{Chew2022, Chew2024}. Figure~\ref{fig2}(d) shows the raw outcome from Ramsey interferometry where the accumulated phase is measured modulo 2$\pi$. We then apply a phase unwrapping algorithm and obtain the intensity profile shown in Fig.~\ref{fig2}(e).
The width of this Gaussian beam measured from this image is ($w_x$, $w_y$) = (556(4), 581(4)) nm, which translates into an effective NA of 0.64 reduced from the one of the objective by residual aberration. Fig.~\ref{fig2}(f) shows a 1D scan for the evaluation of the signal-to-noise ratio (S/N). This technique can obtain as large a signal as possible as long as the phase unwrapping works, i.e., the phase difference between the neighboring points is less than $\pi$. In this measurement, the maximum S/N we obtained is around 170. It is possible to improve the S/N with a smaller step size of the scanning and a better phase-unwrapping algorithm assuming the smooth variation of the light field.

Figure \ref{fig3} shows a gallery of light patterns that we took with the atom camera. Each light pattern is generated by the SLM2 with a different  hologram. We can detect small structures of the light pattern such as a fringe when applying coma abberation on purpose [Fig.~\ref{fig3} (b)]. In Fig.~\ref{fig3}(b), we image an array with a spacing between local maxima of $\sim$ 0.9~{\textmu}m~\cite{Nishimura2024}. The TEM$_{01}$-like pattern shown in Fig.~\ref{fig3}(c) was used in Ref.~\cite{Chew2022} to place two atoms at a distance down to 1.2~{\textmu}m where they experienced ultra-strong Rydberg interaction. 

The spatial resolution of this imaging technique is determined by analyzing the contrast of the image [Fig.~\ref{fig3} (d)]. By deconvoluting the obtained image with a Gaussian PSF (point spread function) with the rms width $\sigma$, we obtained an upper bound of $\sigma \leq 96(4)~{\rm nm}$ (See also the Method in detail). The resolution is better than the one with optical imaging with our objective lens (NA=0.75) and the wavelength for the atomic transition ($\lambda = 780~{\rm nm}$): the rms width of the PSF is 221 nm, thus we achieve the super-resolution. The evaluation of the spatial resolution is limited by the actual contrast of the light pattern itself. Actually, the upper bound of $\sigma$ is considerably larger than the expected rms width of the atomic wavefunction ($\Delta x_{\rm qu} \sim$ 25 nm). The upper bound could be reduced with a pattern with higher intensity contrast.

\subsection*{Polarization profile: Vector light-shift measurement}
%%%%%%%%%%
%%%  Fig 4  %%%
%%%%%%%%%%
\begin{figure}
	\includegraphics[width=\linewidth]{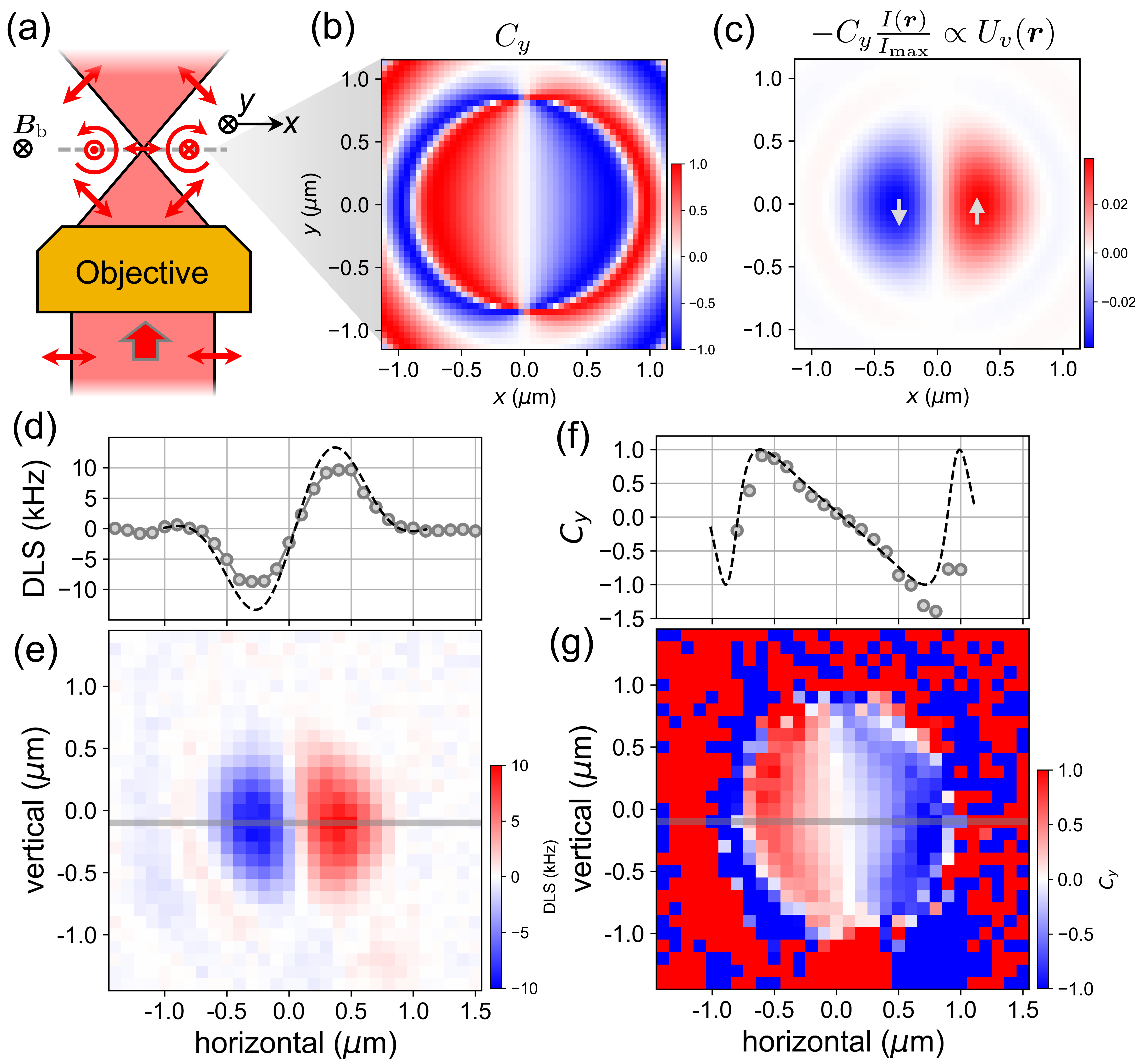}
	\caption{\label{fig4} \textbf{Polarization-imaging mode of the atom camera.} (a) Schematic of the propagation and the polarization of the beam through the high-NA objective lens. The incident light is purely linear polarized light along the $x$ axis. Due to the tight focusing, the light has polarization along the optical ($z$) axis on the focal plane, which gives rise to elliptical polarization rotating clockwise on the right side $(x>0)$ and counter-clockwise on the left side ($x<0$) (b) Numerical simulation of the $y$ component of the ellipticity vector $C_y$ on the focal plane. The calculation is performed with the effective NA of ${\rm NA}_{\rm eff} = 0.64$. (c) Numerical simulation of $-C_y I({\bm r})/I_{\rm max}$, which is proportional to the fictitious magnetic field $ B_{{\rm fict},y}$. The gray arrows indicate the direction of the fictitious magnetic field. (d, e) The 1D and 2D profile of the vector DLS imaged with the stretched transition. The image is shown after the phase unwrapping. The image is tilted due to the mismatch between the scan axis and the polarization axis. (d) The horizontal 1D cut of the image at the center (along the gray line in (e)). The dotted line is a theoretical calculation with the effective NA of ${\rm NA}_{\rm eff}=0.64$ (1D cut of (c)). (f, g) The reconstructed ellipticity $C_y$. (g) The 2D map. The outer part from the center has random values due to the division of the vector DLS by the weak, tail part of the Gaussian intensity distribution. (f) The horizontal 1D cut at the center. Only the data around the center part with reasonable values ($|C_y| \lesssim 1$) are shown. The dotted line is the simulation (1D cut of (b)). 
 }
\end{figure}
We extend this technique to the polarization profile, which, to date, has never been imaged with super-resolution. The polarization is probed using the stretched transition $\ket{2,2}\leftrightarrow\ket{1,1}$ perturbed by the vector DLS. 
%Due to the spin-orbit coupling in the $5P$ states, the light with the circular polarization component (ellipticity) gives rise to polarization-dependent light shift in the $5S$ states called the vector light shift.  
It is helpful to view the vector light-shift as a fictitious magnetic field ${\bm B}_{\rm fict}$  pointing anti-parallel to the ellipticity vector ${\bf C}(\bm r) = {\rm Im}[\bm{\mathcal{E}}({\bm r}) \times \bm{\mathcal{E}}^*({\bm r})]/|\bm{\mathcal{E}}|^2$, where $\bm{\mathcal{E}}({\bm r})$ is the electric field of the light ($|\bm C|=1$ for the purely circularly polarized light and $|\bm C|=0$ for the purely linear polarized light). We also apply a real, much stronger, biased magnetic field ${\bm B}_{\rm b}$ of 3.3 G along the $y$ axis. When ${\bm B}_{\rm fict}$ points also along $y$, then the Zeeman shifts from both fields add and can be measured precisely by probing the magnetic-field sensitive stretched transition using Ramsey interferometry. A simple echo sequence is used to isolate the weak signal (vector LS from the pattern) from the much stronger, and noisy, background field.

As a demonstration, we image the non-trivial polarization profile of an optical tweezers created by tightly focusing a linearly polarized beam [Fig.~\ref{fig4}(a)]. 
When light is tightly focused with a high-NA objective, non-paraxial rays give rise to a longitudinal component of the electric field oscillating along the optical axis~\cite{Richards1959, Boivin1965, Thompson2013, Albrecht2016}. The phase of the longitudinal field is in quadrature with the transverse components, such that the total field has an elliptical polarization in the plane defined by the propagation $z$ axis and the initial polarization along the $x$ axis. The ellipticity vector $\bm{C}$, and the fictitious magnetic field $\bm{B}_{\rm fict}$, are thus oriented along the $y$ axis. Figure~\ref{fig4}(b) shows the theoretical distribution of $C_y$ and (c) shows $-C_y({\bm r}) \times I({\bm r})/I_{\rm max}$ which is proportional to $B_{\rm fict}$. The polarization remains purely linear along the $y$-axis ($x=0$) and becomes elliptical away from it. At the center of the optical tweezers, there is a strong gradient of polarization, proportional to ${\rm NA}^2$, which we calculate to be $dC_y/dx = -1.56$~{\textmu}m$^{-1}$ (see Methods). The corresponding gradient of the vector DLS is 2.05~MHz/{\textmu}m and the gradient of the fictitious magnetic field is $dB_{{\rm fict},y}/dx = 0.98$~G/{\textmu}m for a typical tweezers with $U_s({\bm r=0}) =  -10~{\rm MHz}$, for example.

Figure \ref{fig4} (d, e) shows the measured vector DLS for the tightly-focused Gaussian beam. We clearly observe the characteristic predicted feature of the polarization distribution for the optical tweezers. A 1D cut along the $x$-axis is displayed in Fig.~\ref{fig4}(d) together with the theoretical calculation~\cite{Richards1959} with the effective NA of the system ${\rm NA}_{\rm eff} = 0.64$, which reproduces the beam size obtained from the scalar DLS measurement.
With the intensity profile based on the scalar DLS measurement, we reconstruct ellipticity $C_y$ shown in Figs~\ref{fig4}(f, g). The measured values are well overlapped with the theoretical curve. The gradient of the $C_y$ at the center of the beam is $dC_y/dx = -1.21(2)$~{\textmu}m$^{-1}$, which has a reasonable agreement with the simulation. We suspect that the overestimate arises because the effect of aberration on the polarization gradient is not fully accounted for by the use of an effective NA.

Characterizing the strong gradient of the polarization and the fictitious magnetic field is important in a variety of cold-atom experiments. For example, it strongly perturbs the manipulation of field-sensitive states required for laser cooling~\cite{Thompson2013, Gracia2018}, but it can also be employed as a tool to interrogate the motional state of atoms thanks to the induced spin-motion coupling~\cite{Albrecht2016,Kun-Peng2020,Winkelmann2022}. More generally, the longitudinal components of the electric field in a tightly-focused optical tweezers also affect the high-fidelity qubit control for neutral atom quantum computing with alkaline-earth atoms, which is sensitive to the tensor light shift ~\cite{Unnikrishnan2024} and the trapping of nanoscopic or microscopic objects~\cite{Rohrbach2005, Marago2008, Reece2011}.

\section*{Discussion}

In summary, we have demonstrated the \textit{atom camera}, a technique to image, \textit{in situ}, both the intensity and polarization profiles of light patterns by scanning it with a single ultracold atom probe. The spatial resolution is limited only by quantum fluctuations after cooling the atom to the motional ground-state of the tweezers. We interrogate the local intensity and polarization by using the hyperfine spin degree of freedom of the atom. The excellent coherence property of the hyperfine transition gives rise to a sensitivity an order of magnitude better than in previous schemes based on optical transitions. A Ramsey interferometer including a dynamical decoupling sequence allows us to isolate the signal from the light pattern from other noise sources. We illustrated the technique by imaging various trapping patterns and directly observed, for the first time, the non-trivial polarization profile of a tightly-focused Gaussian beam. 

The camera performance could be improved in several ways. Currently, it takes 40 seconds to take data for a single pixel and hours for a full image. Imaging a large field of view could be performed faster with many atoms probing a pattern in parallel: with a typical atomic distance of $\sim$ 3~{\textmu}m a single atom would have to cover only a 10~{\textmu}m$^2$ area. The duty cycle, currently 20~\%, could also be increased. Here, we used destructive measurement to detect the spin state of the atom such that an atom loading stage (50 ms) is required for each experimental shot. This could be circumvented by implementing non-destructive spin-resolved detection of the atom. Enhanced loading~\cite{Shaw2023, Brown2019, Aliyu2021, Ang2022, Jenkins2022} and atom rearrangement~\cite{Kim2016,Endres2016, Barredo2016} can improve the duty cycle. Another direction for improving the camera performance would be to identify and remove noise sources degrading the atom coherence time (0.1 s) faster than $T_1$ lifetime (several seconds, set by Raman scattering of the tweezers light). 

While we demonstrated the atom camera technique using a pattern at 852 nm, the same wavelength as for trapping the atom, we can image patterns for a wide range of wavelengths with the sensitivity atom camera varying as $|\delta|$ with the detuning from the strong 5$S$-5$P$ transition. The technique can also be extended to image three-dimensional light patterns by scanning the atom along the optical axis using a holographic Fresnel lens~\cite{Barredo2018}. Another application of this sensing method is to precisely measure the depth of the tweezers itself and homogenize it over a large array (currently reaching $< 0.5$~\%)~\cite{Chew2024}, which is important for scaling up neutral atom arrays for quantum simulation and computation~\cite{Schymik2022}. The precise evaluation of the light field profile in a tightly-focused beam by this atom camera will enable one to identify the source of degradation of qubit quality for quantum computers, e.g., crosstalk of local addressing beams~\cite{Radnaev2024} and the non-trivial polarization gradient perturbing the coherence time~\cite{Thompson2013, Unnikrishnan2024}. Finally, instead of using the atom in the quantum ground-state as a probe for the light pattern, our technique could be adapted to image the motional state of the atom~\cite{Drechsler2021,Regal2022} by modifying the Ramsey interferometer~\cite{Alberti2022}.

\section*{Methods}

\subsection{Scalar differential light-shift}

The scalar light shift for the $5S_{1/2}$ state is given by
\begin{equation}
U_s ({\bm r})/h = \frac{\Gamma^2}{8}\frac{I({\bm r})}{I_{\rm sat}}\frac{1}{\delta},
\end{equation}
where $\Gamma$ is the natural linewidth of the $5S-5P$ transition, $I({\bm r})$ is the intensity of the light, $I_{\rm sat}$ is the saturation intensity, and $\delta = -29.6~{\rm THz}$ is the effective detuning from the $5P$ states: $1/\delta = 1/3\delta_{1/2}+2/3\delta_{3/2}$ with the detuning from the $5P_{1/2}$ state $\delta_{1/2}$ and from the $5P_{3/2}$ state $\delta_{3/2}$.
Since the detuning $\delta$ is not exactly the same for $F=1$ and $F=2$ due to the hyperfine splitting $\delta_{\rm HF} = 6.83$ GHz, the amount of the scalar light shift for $F=1, 2$ $U_{s,F=1,2}$ is different, which gives the differential light shift (DLS) $\Delta U_s({\bm r}) = U_{s,F=2}({\bm r}) - U_{s,F=1}({\bm r}) = \eta_s U_s({\bm r})$ with the coefficient $\eta_s = \delta_{\rm HF}/|\delta|$. 
For our tweezers beam at a wavelength of 852 nm, the DLS is 4 orders smaller than the light shift: $\eta_s \simeq \delta_{\rm HF}/|\delta| = 2.31 \times 10^{-4}$. 

\subsection{Vector differential light-shift}

The vector light shift for the $5S_{1/2}$ state is given by ~\cite{Deutsch1998, Corwin1999, Thompson2013}
\begin{equation}
\hat{U}_v({\bm r}) = -U_s({\bm r}) \frac{\delta_{\rm SO}}{2\delta_{1/2}+\delta_{3/2}} {\bf C(\bm r)} \cdot g_F\hat{{\bm F}}. \label{Eq_Vec}
\end{equation}
The direction and the degree of the ellipticity of the light is characterized by the vector ${\bf C}(\bm r) = {\rm Im}[\bm{\mathcal{E}}({\bm r}) \times \bm{\mathcal{E}}^*({\bm r})]/|\bm{\mathcal{E}}|^2$, where $\bm{\mathcal{E}}({\bm r})$ is the electric field of the light ($|\bm C|=1$ for the purely circularly polarized light and $|\bm C|=0$ for the purely linear polarized light). $\hat{\bm F}$ is the total angular momentum operator. Unlike the scalar light-shift, where the dependency of $\hat{\bm F}$ appears implicitly in the difference of the detuning $\delta$, the vector light-shift explicitly depends on $\hat{\bm F}$ due to the spin-orbit coupling. The amount of the vector light-shift $U_v$ can be naturally calculated for a specific $m_F$ state when the $\bf C$ is aligned to the quantization axis. 
For simplicity, the denominator $2\delta_{1/2}+\delta_{3/2}$ in the Eq. (\ref{Eq_Vec}) is approximated as $3\delta$.
The effect of the vector light shift is equivalent to the effect of the Zeeman shift with a fictitious magnetic field ${\bm B}_{\rm fict}$. The direction of the fictitious magnetic field is antiparallel to the direction of the ellipticity vector ${\bm C}$, which is aligned to the $+y$ axis in our setup [Fig.~\ref{fig4} (c)]. In this configuration, the transition frequency is sensitive to the circular polarization rotating around the $y$ axis.

\subsection{Shift of the atom's position}
In this measurement, the potential of the measured pattern should not shift the position of the probe atom. For the atom in the bottom of the tweezers potential with the trap frequency $\omega$, the potential gradient $k$ generated by the target light pattern leads to the displacement of the position of the probe atom $\Delta x_{\rm shift} = k/m\omega^2$, which results in the degradation of the resolution. To benefit from the advantage of having cooled the atom to the motional ground state, the atom's position shift $\Delta x_{\rm shift}$ should not be significantly larger than the quantum uncertainty of the atom's position $\Delta x_{\rm qu} = \sqrt{\hbar/2m\omega}$, which is the rms uncertainty of the probability distribution. These values are comparable when $k \sim \hbar\omega/2\Delta x_{\rm qu} = h \times 2~{\rm kHz}/{\rm nm}$ for the parameters in our setup. When the target light pattern is a Gaussian beam with the same waist as the tweezers beam, we ensure $\Delta x_{\rm shift} < \Delta x_{\rm qu}$ by keeping the target light peak intensity at 10\% of the trap intensity, giving a force with largest slope of the potential of $h \times 2~{\rm kHz}/{\rm nm}$.

\subsection{Ramsey interrogation}

We perform Ramsey interferometry to measure the DLS. To detect the weak signal of the target light pattern without being perturbed by the noisy environment, we apply a dynamical decoupling sequence in the Ramsey interferometry. We insert an XY-4 sequence for the scalar DLS measurement whereas we use a simpler echo sequence for the vector DLS measurement with a relatively larger signal. We start the Ramsey interrogation by applying a 6.8-GHz microwave $\pi/2$-pulse and then apply 4 $\pi$-pulses with alternating phase between $X$ and $Y$ [Fig.~\ref{fig2}(a)]. During the 4 pulses, the target light is toggled accordingly so that its DLS signal remains while removing the (large and noisy) contribution from the trap. The target light is linearly ramped up and down in 1 ms to minimize the heating of the probe atom. The total phase accumulation $\phi$ is $\phi = \int \Delta U(t)/h ~dt = \Delta U/h \times \tau_{\rm eff}$.

To interrogate the weak scalar DLS from the pattern ($\sim 100$s Hz), we choose a total Ramsey time of 40~ms (the effective interrogation time $\tau_{\rm eff} = 16$~ms is shorter because of the decoupling sequence and the adiabatic ramping of the pattern to avoid heating the atom). The Ramsey interferometer is completed well within the coherence time of the clock transition $T_2^{\rm XY4} > 100~{\rm ms}$. The imperfect fringe visibility $V \sim 80$~\% is mainly caused by preparation and detection errors. In Fig.~\ref{fig2}(b), we finely sample the Ramsey fringe for illustration, however this is not required to estimate the fringe phase. To minimize data acquisition time when recording full 2D image of the pattern, we only acquire data for a closing $\pi/2$-pulse phase of 0$^{\circ}$ and 90$^{\circ}$, which is the minimal dataset required. We repeat the experiment 100 times ($N \sim$ 50 shots with atoms due to the $\sim$ 50~\% loading probability), with a cycling time of 200 ms, to estimate the probability for the atom to be in the $F=1$ state. 

\subsection{Spatial resolution} 

The image profile obtained by this imaging technique is the convolution of the spatial profile of the actual target light and the spatial distribution of the probe (point spread function). By measuring a light profile with a sharp variation, such as the one around the center of the TEM$_{01}$-like profile in Fig.~\ref{fig3}(c,d), we can evaluate the resolution. For simplicity, we consider the 1D system. We assume a target light profile with a quadratic shape
\begin{equation}
I(x)=ax^2+b
\end{equation}
and the atom camera has a point spread function with a Gaussian distribution:
\begin{equation}
\rho(x) = \rho_0 e^{-\frac{(x-x_0)^2}{2\sigma^2}}.
\end{equation}
Here $x_0$ is the center of the probe atom, $\rho_0$ is the normalization constant, and $\sigma$ is the root-mean-square width, which characterizes the resolution. The DLS $\Delta U(x_0)$ obtained by scanning the center of the atom $x_0$ is proportional to the convolution of $\rho(x)$ and $I(x)$ 
\begin{align}
\Delta U(x_0) &\propto \int dx \rho(x-x_0)I(x) \\
&= ax_0^2 + a\sigma^2 + b,
\end{align}
which shows that the offset of the DLS measured at the center $x_0=0$ ($a\sigma^2 + b$) is larger than the original offset $b$ by $a\sigma^2$ due to the finite resolution $\sigma$ while the measured curvature (the quadratic coefficient $a$) is independent of the resolution $\sigma$. Since the measured offset $a\sigma^2 + b$ is never below this additional offset $a\sigma^2$ ($a\sigma^2 \leq a\sigma^2 + b$), we can put the upper bound of $\sigma^2$:
\begin{equation}
\sigma^2 \leq \frac{a\sigma^2 + b}{a}.
\end{equation}

We measured the scalar DLS for the 1D cut of the TEM$_{01}$-like pattern [Fig. \ref{fig3} (d)]. The phase shift in the Ramsey spectroscopy is fitted with a quadratic function to determine the quadratic coefficient $a$ and the offset $a\sigma^2 + b$, even though the actual values of $\sigma$ and $b$ remain unknown. From these values, we obtained the upper bound of the rms width $\sigma \leq 96(4)~{\rm nm}$.

\subsection{Non-paraxial tweezers}
From the vector diffraction theory~\cite{Richards1959}, we can derive the theoretical value of the polarization gradient at the center for a Gaussian illumination. For our choice of a Gaussian beam illumination with a $1/e^2$ width 2.34 times larger than the radius of the lens aperture, the gradient of the ellipticity can be approximated by $dC_y/dx \simeq 3.02\alpha\sin{\alpha}/\lambda$, where $\alpha=\sin^{-1}{\rm NA}$. % (generalize?). 
From the intensity profile obtained by the image of the scalar DLS, we can define the effective numerical aperture ${\rm NA}_{\rm eff}$, which reproduces the actual beam width from the formula. In our system, ${\rm NA}_{\rm eff} = 0.64$ while the NA in the design is NA = 0.75. With ${\rm NA}_{\rm eff}$ and $\lambda=852~{\rm nm}$, the gradient of the ellipticity is $dC_y/dx = -1.56$~{\textmu}m$^{-1}$.

\section*{}
\begin{acknowledgements}
\textbf{Acknowledgements} This work was supported by MEXT Quantum Leap Flagship Program (MEXT Q-LEAP) JPMXS0118069021, JSPS Grant-in-Aid for Specially Promoted Research Grant No. 16H06289, JSPS Grant-in-Aid for Research Activity Start-up No. 19K23431, JSPS Grant-in-Aid for Transformative Research Areas No. 22H05267, and JST Moonshot R\&D Program Grant Number JPMJMS2269.

\textbf{Author contributions} T.T., Y.T.C., R.V., T.P.M., and S.d.L. designed and carried out the experiments. H.S., K.N., and T.A. developed the holograms for the target light patterns and contributed to designing and constructing the optical system. T.T. performed data analysis, theory, and simulation. T.T. and S.d.L. wrote the manuscript with contributions from all authors. S.d.L. and K.O. supervised and guided this work.

\end{acknowledgements}

%\bibliographystyle{apsrev4-2}
%\bibliography{abbreviated}
%\bibliography{bibfile_AtomCamera}

\providecommand{\noopsort}[1]{}\providecommand{\singleletter}[1]{#1}%

\end{document}